# Conceptual quantification of the dynamicity of longitudinal social networks

Shahadat Uddin, Mahendra Piraveenan, Arif Khan and Babak Amiri
Complex Systems Research Centre, Faculty of Engineering & IT
The University of Sydney
Sydney, Australia
{shahadat.uddin, mahendrarajah.piraveenan, mkha6767, amiri.babak}@sydney.edu.au

*Abstract*—A longitudinal social network evolves over time through the creation and/or deletion of links among a set of actors (e.g. individuals or organisations). Longitudinal social networks are studied by *network science* and *social science* researchers to understand network evolution, trend propagation, friendship and belief formation, diffusion of innovations, the spread of deviant behaviour and more. In the current literature, there are different approaches and methods (e.g. Sampson's approach and the Markov model) to study the dynamics of longitudinal social networks. These approaches and methods have mainly been utilised to explore evolutionary changes of longitudinal social networks from one state to another and to explain the underlying reasons for these changes. However, they cannot quantify the level of dynamicity of the over time network changes and the contribution of individual network members (i.e. actors) to these changes. In this study, we first develop a set of measures to quantify different aspects of the dynamicity of a longitudinal social network. We then apply these measures, in order to conduct empirical investigations, to two different longitudinal social networks. Finally, we discuss the implications of the application of these measures and possible future research directions of this study.

## I. Introduction

The study of longitudinal social networks has been the subject of intense research interest in recent years [1, 2] because it provides a way to analyse the underlying mechanism in the process of network formation, development and evolution over time [3]. Researchers have been exploring longitudinal social networks in order to understand a wide range of processes in various contexts, such as knowledge creation in co-authorship networks [4, 5], spread of virus in computer networks [6] and spread of happiness and obesity in kinship networks [7, 8]. Organisations are also interested in studying longitudinal network in order to get inside the decision cycle of major events [9]. There is a growing interest in studying longitudinal social networks in other research areas, for example, education, psychology, health study, childhood and youth study, life history, organisation science and criminology [10-12].

In *network science* and *social science* literature, the presence of methods and approaches for the analysis of longitudinal social networks has been noticed for quite some time. One of the most notable and earliest approaches to the study of dynamics of longitudinal social network is the Sampson's [13] approach that he followed in his monastery study. In this study of the dynamics of a longitudinal social network, he took snapshots of the same network from different intervals, and observed and analysed the evolution of the network. The other dominant methods for analysing longitudinal social networks are Markov models and Multi-agent simulation models. Continuous time Markov chains for modelling longitudinal social networks were proposed as early as 1977 by Holland and Leinhardt [14], which have been significantly improved later by many researchers [15-18]. The most important property of a Markov model is that the future state of a process is dependent only on the current state but not on any previous state [19]. For modelling longitudinal social networks, exponential random graph and stochastic actor-oriented models are the two Markovian methods proposed by Robins et al. [17] and Snijders et al. [20] respectively. In these two approaches of network analysis, links are modeled as random variables that can be in different states (e.g. positive, negative or neutral) at different time. The purpose of this link-modelling approach is to examine which network effect fits the empirical data and better accounts for the observed structural changes. These two Markovian approaches to longitudinal social network analysis are efficient enough to detect and describe network changes and to explain why these changes occur. However, they may have convergence issues in the presence of sufficiently large abrupt endogenous (i.e. structure based) and exogenous (i.e. attribute based) social changes [21]. In the Multi-agent simulation model, members in a social network are often modeled and implemented as computer agents who have the abilities to behave and make decisions based on certain criteria. The collective behaviours' of all members in a network will determine how the network evolves from one structure to another. Evolutionary models often use multi-agent simulation. Carley et al. [22] use multi-agent technology to simulate the evolution of covert networks such as terrorist groups. Moreover, using a multi-agent system called DYNET they perform a '*what-if*' analysis to anticipate how a network adapts to environmental changes such as the removal of a central member. A simulation model can be a powerful tool for predicting a network's future. However it often oversimplifies the behavior and decision-making of humans, and may not be able to model the complex reality of social networks [23]. Like Sampson's approach, the Markov and Multi-agent simulation models are also unable to quantify the level of dynamic behaviour shown by an individual network member or a group of network members in any longitudinal setting [24].

The methods and approaches for exploring longitudinal social networks available in the present literature give emphasis mainly to the holistic view of network for studying network dynamics and are therefore unable to quantify different aspects of the dynamicity of a longitudinal social network. This limitation further hinders the introduction of an effective approach to compare and contrast two (or more) different longitudinal social networks [11]. This study aims to overcome this shortcoming by proposing a set of measures to quantify different aspects of network dynamicity of a given longitudinal social network. The rest of this contribution is organised as follows. In section two, we construct a set of measures for quantifying different aspects of the dynamicity of a given longitudinal social network. These measures are utilised to explore two real longitudinal social networks in section three. Section four discusses the contribution of this study to the present literature. Finally, there is a conclusion and an illustration of possible future research directions in section five.

## II. CONSTRUCTING MEASURES FOR LONGITUDINAL SOCIAL NETWORK

Longitudinal social networks are being observed at different time points to collect network data for research analysis purposes. These observed networks are named as *short-interval* networks. The accumulation of these *short-interval* networks creates another bigger network, which is termed as an *aggregated* network. Based on the concept of *static* and *dynamic* social network topology, this study develops a set of measures to quantify different aspects of the dynamicity of longitudinal social networks. Social network topology defines the way that a given longitudinal social network will be analysed in terms of over time aggregation of links among network members [11, 25]. In *static topology*, methods of social network analysis (SNA) are applied on the *aggregated* network of an entire data collection period; whereas, smaller segments of network data accumulated in less time compared to the entire data collection period are used in *dynamic topology* for research analysis purposes [25, 26]. For instance, a *dynamic topology* could be exercised on daily or weekly or even monthly network of a university email communication network that evolves over five years, while *static topology* considers only one network - the single *aggregated* network of five years. Figure 1 shows a schematic difference between these two types of SNA topologies. In this figure, two longitudinal social networks, that are observed in three points of time (i.e. *Day1*, *Day2* and *Day3*), evolve over time. According to this figure, for *static* network analysis purposes SNA methods are applied to the *aggregated* network (i.e. the upper shaded network inside the square of the first longitudinal social network) at the end of *Day3*. In contrast, SNA methods are applied to each day network for research analysis purposes in *dynamic topology* (i.e. the three lower shaded networks inside squares of the second longitudinal social network). There is no *aggregated* network considered for network analysis in this topology. That means *dynamic topology* explores structural positions of actors in different sets of network data that are collected in a shorter time period compared to the overall duration of the longitudinal social network. The *static topology* explores only one network which is constructed by aggregating all links and actors of different sets of network data utilised in the *dynamic topology*.

The level of dynamicity shown by a longitudinal social network relies on the changes of positional behaviours (e.g. degree centrality and closeness centrality) of actors in all *short-interval* networks compared to the *aggregated* network. In order to explore the dynamicity of a longitudinal social network, it is therefore required to observe and analyse the involvements of individual actors (i) in all *short-interval* networks; and (ii) in the *aggregated* network. To capture dynamics of networks that emerge in *short-interval* networks, the *dynamic topology* needs to be followed. On the other hand, *static topology* has to be carried out for the single *aggregated* network. Thus, to explore longitudinal social networks, both *static* and *dynamic* topological analyses of networks need to be carried out.

The structural positions of an individual actor in *short-interval* networks of a longitudinal social network represent the pattern of the changes of its network behaviour. This can further reveal how actors change their network roles (e.g. network positions and level of interactions with other actors) in *short-interval* networks over time. The structural positions of individual actors of a longitudinal social network can be calculated by using basic social network analysis measures (e.g. degree centrality, closeness centrality and network constraint). This study defines the *degree of dynamicity* (or *level of dynamicity* or simply *dynamicity*) shown by an individual actor as the variability of the structural positions of that actor in all *short-interval* networks compared to its structural position in the *aggregated* network. The following equation represents the *degree of dynamicity* (or dynamicity) shown by an individual actor in a longitudinal social network:

$$DDA^i = \frac{\sum_{j=1}^{m} | OV^i_{AN} - OV^i_{SIN(j)} |}{m} \quad\quad\quad\quad (1)$$

Where, $DDA^i$ represents *degree of dynamicity* shown by the $i^{th}$ actor, $OV^i_{AN}$ indicates observed value (say degree centrality) in the *aggregated* network for the $i^{th}$ actor, $OV^i_{SIN(j)}$ indicates observed value for the same SNA measure (i.e. degree centrality) in the $j^{th}$ short-interval network (SIN) for the $i^{th}$ actor and $m$ indicates the number of short-interval networks considered in the analysis. Since the scaled value of any network attribute (e.g. degree centrality) for an actor in a social network will have the range between 0 and 1 [27], $OV_{AN}^i$ and $OV_{SIN(j)}^i$ have the range between 0 and 1. Therefore, the range for $DDA^i$ is between 0 and 1.

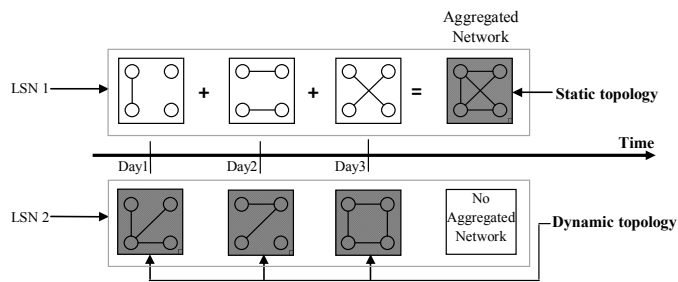

Fig 1. Illustration of static and dynamic topology of social network analysis. LSN stands for Longitudinal Social Network

TABLE 1. POSSIBLE COMBINATION OF PRESENCE AND ABSENCE OF AN ACTOR IN TWO CONSECUTIVE SHORT-INTERVAL NETWORKS

| Current SIN (Present/Absent) | Previous SIN (Present/Absent) | $\alpha_{i,i-1}$ |
|---|---|---|
| Present | Present | $\alpha_{p,p} = 1.0$ |
| Present | Absent | $\alpha_{p,a} = 0.5$ |
| Absent | Present | $\alpha_{a,p} = 0.0$ |
| Absent | Absent | $\alpha_{a,a} = 0.0$ |

In a given longitudinal social network, an actor may not be found in all short-interval networks. An actor may participate in the $j^{th}$ short-interval network; however, it may not participate in the $(j-1)^{th}$ short-interval network. Or, it could be the case that an actor is present in the current short-interval network but will be absent in the subsequent short-interval network. The possible combination of 'presence' and 'absence' of an actor in two consecutive short-interval networks is illustrated in Table 1. When an actor is absent in the $j^{th}$ short-interval network, the value of $OV_{SIN(j)}^i$ in Equation 1 for that actor will be zero. In terms of 'presence' and 'absence' in two consecutive short-interval networks, an actor who is present in the $j^{th}$ and $(j-1)^{th}$ short-interval networks will show higher level of dynamicity compared to any other actor who is present in the $j^{th}$ short-interval network but absent in the $(j-1)^{th}$ short-interval network. That means a transition from the 'absence' state in the $(j-1)^{th}$ short-interval network to the 'presence' state in the $j^{th}$ short-interval network will negatively impact the shown dynamicity for an actor. In order to capture the contribution of this type of phase transition to the shown dynamicity, a constant is introduced to the Equation 1:

$$DDA^i = \frac{\sum_{j=1}^{m} \alpha_{j,j-1} \times |OV_{AN}^i - OV_{SIN(j)}^i|}{m} \quad \ldots (2)$$

The values of $\alpha_{j,j-1}$ for all the possible combinations of 'presence' state and 'absence' state of an actor in two consecutive short-interval networks are presented in Table 1. When there is a phase transition from the 'absence' state to the 'presence' state in two consecutive short-interval networks for an actor, $\alpha_{j,j-1}$ will be 0.5. If an actor is absent in the present short-interval network then $\alpha_{j,j-1}$ will be 0 and it will be 1 when an actor is present in two consecutive short-interval networks. For the first short-interval network (i.e. $\alpha_{1,0}$ for j=0) it will be 1. Since the maximum value of $\alpha_{j,j-1}$ can be 1, the range for $DDA^i$ in Equation 2 will be between 0 and 1.

Equation 2 can quantify the level of dynamicity shown by individual actors in a given longitudinal social network regardless of its size and the number of short-interval networks constitutes the aggregated network. It can be conceptualised that the dynamicity shown by a longitudinal social network is the reflection of the dynamicities represented by its all member actors. The dynamicity shown by each actor is normalised using the highest observed dynamicity for an actor in the longitudinal social network. Therefore, the contribution of individual actors to the dynamicity of the longitudinal social network can be quantified by the following equation:

$$DDN^i = \frac{1 - (DDA^* - DDA^i)}{n} \quad \ldots (3)$$

Where, $DDN^i$ represents the contribution of the $i^{th}$ actor to the degree of dynamicity shown by the longitudinal social network, $DDA^*$ is the highest observed degree of dynamicity shown by individual actors in the network, $DDA^i$ is the degree of dynamicity for the actor $i$ and $n$ is the number of actors in the network. Since the range for $DDA^*$ and $DDA^i$ is between 0 and 1, the range for $DDN^i$ is between 0 and $1/n$.

To calculate the degree of dynamicity shown by a short-interval network, it is required to compare the network position of all actors of that short-interval network with their positions in the aggregated network. This is represented by the following equation:

$$DDN^{SIN(i)} = \frac{\sum_{j=1}^{w_i} \alpha_{i,i-1}^j \times |OV_{AN}^j - OV_{SIN(i)}^j|}{w_i} \quad \ldots (4)$$

Where, $DDN^{SIN(i)}$ represents the dynamicity shown by the $i^{th}$ short-interval, $\alpha_{i,i-1}^j$ represents $\alpha_{i,i-1}$ (same as in the Equation 2) for actor $j$, $OV_{AN}^j$ indicates the observed value (say degree centrality) of the $j^{th}$ actor in the aggregated network and $OV_{SIN(i)}^j$ indicates the observed value of the $j^{th}$ actor in the $i^{th}$ short-interval network and $w_i$ is the total number of actors in the $i^{th}$ short-interval network. Precisely, $DDN^{SIN(i)}$ indicates the average dynamicity shown by an actor of the $i^{th}$ short-interval. The range for $DDN^{SIN(i)}$ will be between 0 and 1 since the range for $OV_{AN}^j$ and $OV_{SIN(i)}^j$ is between 0 and 1.

In order to calculate the degree of dynamicity shown by the longitudinal social network, the right hand side of Equation (3) needs to be summed up for all actors. Therefore,

$$DDN = \sum_{i=1}^{n} DDN^i \quad \ldots (5)$$

Using Equation (3) to replace $DDN^i$ in Equation (5), we will get

$$DDN = \frac{\sum_{i=1}^{n} [1 - (DDA^* - DDA^i)]}{n} \quad \ldots (6)$$

Where, DDN represents degree of dynamicity shown by the longitudinal social network, $DDA^*$ is the highest observed degree of dynamicity shown by an individual actor in the network, $DDA^i$ is the *degree of dynamicity* for actor $i$ and $n$ is the number of actors in the longitudinal network.

TABLE 2. TOP-5 ACTORS SHOWING HIGHER DYNAMICITY ($DDA^i$) BASED ON BASIC CENTRALITY MEASURES FOR ORGANISATIONAL COMMUNICATION NETWORK

| Degree Centrality | | Closeness Centrality | | Betweenness Centrality | |
|---|---|---|---|---|---|
| Actor ID | Dynamicity | Actor ID | Dynamicity | Actor ID | Dynamicity |
| 71 | 0.00385 | 157 | 0.3600 | 13 | 0.07110 |
| 85 | 0.00337 | 129 | 0.3368 | 110 | 0.06392 |
| 128 | 0.00306 | 450 | 0.3321 | 43 | 0.05517 |
| 13 | 0.00248 | 132 | 0.3197 | 20 | 0.05242 |
| 20 | 0.00226 | 199 | 0.3087 | 35 | 0.04141 |

TABLE 4. *DEGREE OF DYNAMICITY*, IN REGARDS TO THREE BASIC CENTRALITY MEASURES, SHOWN BY TWO DIFFERENT LONGITUDINAL SOCIAL NETWORKS

| Longitudinal Social Networks | Dynamicity shown by the longitudinal social network | | |
|---|---|---|---|
| | Degree | Closeness | Betweenness |
| Enron network | 0.003807 | 0.338271 | 0.069986 |
| Students' email network | 0.083991 | 0.445498 | 0.196822 |

### III. APPLICATION OF THE PROPOSED MEASURES

The measures constructed in the previous section can quantify dynamicity of a longitudinal social network at different levels. In this section, we explore two longitudinal social networks using the measures represented in Equation 2, Equation 4 and Equation 6. The first longitudinal social network is considered from the context of organisational communication network and the second one consists of students' email communications that evolve throughout a university semester.

#### A. Organisational Communication Network

The email communication dataset from Enron, commonly referred as Enron email corpus, has been analysed using the proposed measures in this example. This corpus was released by Federal Energy Regulatory Commission (FERC) in May, 2002. Shetty and Adibi [28] from University of Southern California created a MySQL database of this corpus. They also cleaned the database by removing a large number of duplicate emails, computer generated folders, junk data and invalid email addresses. In the area of organisational science and social networking research, the Enron corpus is of great value because it allows the academic to conduct research on real-life organisation over a number of years. It is well documented in the literature that a drastic form of critical loss, which was being started to flourish in the beginning of the third quarter of 2001, occurs in Enron during the final quarter of 2001 [29]. In this empirical example, we consider a portion of the email communications of Enron which evolve during the second half of the year 2001 (i.e. from July to December 2001). This portion of the Enron dataset contains 735261 messages from 2253 distinctive users. A *short-interval* network consists of email communications that evolve during a month. Therefore, there are six *short-interval* networks and one *aggregated* network considered for research data analysis in this example.

TABLE 3. *DYNAMICITY* SHOWN BY *SHORT-INTERVAL* NETWORK ($DDN^{SIN_i}$) OF THE ORGANISATIONAL COMMUNICATION NETWORK

| SIN ID | Dynamicity based on basic centrality measures | | |
|---|---|---|---|
| | Degree | Closeness | Betweenness |
| 1 | 0.000049 | 0.02696 | 0.001564 |
| 2 | 0.000046 | 0.03146 | 0.001549 |
| 3 | 0.000050 | 0.02962 | 0.001578 |
| 4 | 0.000086 | 0.06254 | 0.002216 |
| 5 | 0.000147 | 0.06216 | 0.002024 |
| 6 | 0.000072 | 0.03721 | 0.001897 |

Table 2 presents the top 5 actors of the Enron's email network in terms of *dynamicity* (i.e. $DDA^i$). In calculating these *dynamicity* values, we consider only three actor-level network centralities (i.e. degree, closeness and betweenness). That means this table presents top 5 actors that show higher *dynamicity* in terms of degree centrality, closeness centrality and betweenness centrality in the Enron's longitudinal network. The dynamicities, in terms of all three basic centrality measures, shown by each of the six *short-interval* networks (i.e. $DDN^{SIN(i)}$) for Enron email dataset are presented in Table 3. In regards to basic centrality measures, *the level of dynamicity* (i.e. DDN) shown by the Enron's longitudinal social network is presented in the second last row of Table 4.

There is an overlapping of actors' positions in the top-ranked lists of *degree-dynamicity* and *betweenness-dynamicity*. Two actors with ID 13 and 20 are found (see Table 2) in the lists of top 5 actors showing higher *degree-dynamicity* and *betweenness-dynamicity*. Dynamicities shown by fourth and fifth *short-interval* networks in respect of all three centrality measures are higher, as noted in Table 3, compared to the other *short-interval* networks (i.e. first, second, third and sixth *short-interval* network). It is well documented in the literature that the organisational crisis of Enron was at its peak during this period (i.e. October and November 2001) which resulted in the bankruptcy declaration during the first week of December 2001 [30]. Therefore, the measures proposed in this study are able to explore the underlying external influences (e.g. organisational crisis of Enron) to the different phases (i.e. *short-interval* networks) of the longitudinal social network.

#### B. Students' Communication Network

In this example, we utilise a students' email communication network dataset. This communication network was evolved among 34 students during a university semester consisting of 3 months. These 34 students were doing a masters-degree course, entitled *Statistical Methods in Project Management*, which was delivered in *face-to-face* mode. For all course-related communication, students were motivated and advised to communicate with other students as well as with the tutor and the lecturer of the course only through a designated email communication system. Those emails that have a single recipient are considered for research analysis as this type of emails reflect more intensive and directed communications [31]. After conducting all required refinements, 621 emails were found in the research dataset. Three *short-interval* networks and an *aggregated* network are considered for longitudinal data analysis.

Table 5 presents top 5 actors that show higher *dynamicity* (i.e. $DDA^i$) in terms of three basic centrality measures in the students' email communication network. The dynamicities shown by each of three *short-interval* networks (i.e. $DDN^{SIN(i)}$) of the students' email communications are presented in Table 6. The *dynamicity* (i.e. DDN) shown by the longitudinal students' email network is presented in the last row of Table 4.

Three actors (i.e. actor with ID 2, 32 and 4) are found in the top-ranked lists of *degree-dynamicity* and *betweenness-dynamicity*. *Degree-dynamicity* shows an increasing pattern among three *short-interval* networks. With the increase of study load throughout a semester, students make more email communication with their peers [32], which will eventually lead to an increased *degree-dynamicity* shown by the different *short-interval* networks.

## IV. DISCUSSION

Based on the concept of *static* and *dynamic* network topology, in this study we develop a set of measures to explore dynamicity of a longitudinal social network. We then utilise these measures to explore two longitudinal social networks (i.e. organisational communication network and students' communication network). For these two longitudinal social networks, we show: (i) top 5 actors that reveal higher *dynamicity* as captured by the Equation 2; (ii) the *dynamicity* shown by each of the *short-interval* networks as quantified by the Equation 4; and (iii) the *dynamicity* shown by the longitudinal social network, which has been calculated by the Equation 6.

The proposed measures of this study are able to explore *dynamicity* of a given longitudinal social network in respect of any actor-level network attribute. Although we only consider basic centrality measures (i.e. degree centrality, closeness centrality and betweenness centrality) for the empirical study of two longitudinal social networks using the proposed measures, other actor-level social network attributes (e.g. information centrality, in-degree and out-degree) can be considered. This will further enable researchers to explore the *dynamicity* of a given longitudinal social network from a wide range of perspectives. For instance, researchers can utilise the *in-degree* attribute in the proposed measures of this study to explore *dynamicity* of the activity of actors in a given longitudinal social network since the *in-degree* represents activity of actors in a given social network [27]. Similarly, *out-degree* can be used in the proposed measures of this study to explore *dynamicity* of the popularity of actors in a given longitudinal social network.

Existing methods (e.g. Markov model) of current literature for analysing longitudinal social networks are unable to quantify the contribution of an individual actor to the overall evolutionary dynamicity of a given longitudinal social network [21]. The measure proposed in the Equation 3 is able to overcome this shortcoming. Using this measure, researchers now can explore involvements of actors (e.g. which actor is playing major in the network development process) throughout the evolution of a given longitudinal social network. Moreover, this study proposes another measure in the Equation 6 to calculate the level of dynamicity shown by a longitudinal social network. This measure eventually enables researchers to compare the network-level *dynamicity* of two or more longitudinal social networks regardless of their network sizes, the number of interactions among their member actors and the number of the *short-interval* networks constitutes the *aggregate* network.

The measure proposed in the Equation 4 can determine the *dynamicity* shown by a *short-interval* network. It can further reveal the network statistics of the corresponding *short-interval* network. For example, if the measure proposed in the Equation 4 shows a low value for the second *short-interval* network of a given longitudinal social network then it can concluded that (i) a lower number of actors participate in *that short-interval* network compared to other *short-interval* networks and the *aggregated* network; and/or (ii) most of actors participated in that *short-interval* do not participate in the first *short-interval* network ($\alpha_{j,j-1}$ will be 0.5 in that case; thus, lowering the numerical value of the Equation 4).

## V. CONCLUSION AND FUTURE RESEARCH

In the present literature, there are many studies that propose methods and approaches to explore longitudinal social networks. Those studies mostly give emphasis to explore the underlying process for network development and evolution. This study takes the initiative to quantify different aspects of the *dynamicity* of a longitudinal social network. The proposed measures of this study will create opportunities for researchers to explore a given longitudinal social network from different perspectives (e.g. which actor contributes more to the evolution of a longitudinal social network and the *dynamicity* shown by a *short-interval* network).

TABLE 5. TOP-5 ACTORS SHOWING HIGHER DYNAMICITY ($DDA^i$) BASED ON BASIC CENTRALITY MEASURES FOR STUDENTS' COMMUNICATION NETWORK

| Degree Centrality | | Closeness Centrality | | Betweenness Centrality | |
|---|---|---|---|---|---|
| Actor ID | Dynamicity | Actor ID | Dynamicity | Actor ID | Dynamicity |
| 2 | 0.10875 | 10 | 0.65227 | 32 | 0.23069 |
| 32 | 0.09239 | 23 | 0.56904 | 2 | 0.18307 |
| 23 | 0.06397 | 3 | 0.55960 | 7 | 0.12922 |
| 6 | 0.05290 | 34 | 0.41451 | 9 | 0.09728 |
| 4 | 0.04965 | 1 | 0.40255 | 4 | 0.06982 |

TABLE 6. *DYNAMICITY* OF *SHORT-INTERVAL* NETWORK ($DDN^{SIN(i)}$) OF THE STUDENTS' COMMUNICATION NETWORK

| SIN ID | Dynamicity based on basic centrality measures | | |
|---|---|---|---|
| | Degree centrality | Closeness centrality | Betweenness centrality |
| 1 | 0.011903 | 0.08460 | 0.01083 |
| 2 | 0.012069 | 0.13616 | 0.02373 |
| 3 | 0.015999 | 0.07307 | 0.01868 |

This study creates future research opportunities in a number of ways. First, we can explore the *dynamicity* shown by individual actors in each *short-interval* network using a variation of the Equation 2 (i.e. without the summation operator). It will further enable to study the changing network behaviour of actors in a longitudinal setting. Second, to capture the *dynamicity* due to a phase change (i.e. from the '*presence*' state to the '*absence*' state) of an actor, this study introduced a constant (i.e. $\alpha_{j,j-1}$). The possible values that this constant can take (i.e. 0, 0.5 and 1) for any longitudinal social network are presented in Table 1. This value assignment may not capture the appropriate quantity of a phase-change *dynamicity* shown by casual or part-time network members. In an organisation, for example, a part-time employee works on Monday, Wednesday and Friday. If *short-interval* networks consist of a *day* then she will be found to be responsible for many phase changes in the email communication network of that organisation although she participates in the communication network during her regular office hour. Further research investigations in regards to actors' network membership and connectivity are required in order to assign correct values to this constant for different types of phase changes. Finally, the pattern of over time interaction among actors can be examined by using the measures that capture the *dynamicity* shown by an actor and the contribution of individual actors to the *dynamicity* shown by the longitudinal social network (i.e. measures represented by Equation 2 and Equation 3 respectively). For instance, these measures can be utilised to explore whether there is a tendency of link establishments in subsequent *short-interval* networks among actors that are highly connected with other actors at the present *short-interval* network.